\documentclass[12pt]{article}
\usepackage{color}

\newcommand{\beq}{\begin{equation}}
\newcommand{\eeq}{\end{equation}}
\newcommand{\beqa}{\begin{eqnarray}}
\newcommand{\eeqa}{\end{eqnarray}}
\newcommand{\beqar}{\begin{eqnarray*}}
\newcommand{\eeqar}{\end{eqnarray*}}

\newcommand{\al}{\alpha}
\newcommand{\be}{\beta}

\def\Tr           {\mbox{\rm Tr}\,}

\def\ran          {\rangle}
\def\lan          {\langle}
\def\fsC    {C\!\!\!\!/\,}
\def\fsH    {H\!\!\!\!/\,}

\newcommand{\eps}{\epsilon}

\newcommand{\lam}{\lambda}

\newcommand{\labell}[1]{\label{#1}} 
\newcommand{\reef}[1]{(\ref{#1})}
\newcommand\prt{\partial}

\def\sst#1{{\scriptscriptstyle #1}}
\def\0{{\sst{(0)}}}
\def\1{{\sst{(1)}}}
\def\2{{\sst{(2)}}}
\def\3{{\sst{(3)}}}
\def\4{{\sst{(4)}}}
\def\5{{\sst{(5)}}}
\def\6{{\sst{(6)}}}
\def\7{{\sst{(7)}}}
\def\8{{\sst{(8)}}}

\begin{document}
\baselineskip 18pt%
\begin{titlepage}
\vspace*{1mm}%
\hfill
TUW-16-09
\vbox{

    \halign{#\hfil         \cr
           } 
      }  
\vspace*{8mm}
\vspace*{8mm}%

\center{ {\bf \Large  
 	
On  Bulk Singularity Structures and all order $\alpha'$ Contact  Terms of  BPS String Amplitudes  }} 

\begin{center}
{Ehsan Hatefi   }\footnote{
ehsan.hatefi@tuwien.ac.at,ehsan.hatefi@cern.ch, e.hatefi@qmul.ac.uk}

\vspace*{0.6cm}{ Institute for Theoretical Physics, TU Wien
\\
Wiedner Hauptstrasse 8-10/136, A-1040 Vienna, Austria}
\vskip.06in

\vspace*{.3cm}
\end{center}
\begin{center}{\bf Abstract}\end{center}
\begin{quote}

 The entire form of the amplitude of  three SYM ( involving two transverse  scalar fields, a gauge field)  and 
 a potential  $C_{n-1}$ Ramond-Ramond  (RR) form field  is found out. We first derive  $<V_{C^{-2}}  V_{A^{0}} V_{\phi ^{0}}  V_{\phi ^{0}}>$ and then 
 start constructing an infinite number of $t,s$ channel bulk singularity structures by means of all order $\alpha'$ corrections to
pull-back of brane in an Effective Field Theory (EFT). Due to presence of the complete form of S-matrix, several new contact interactions 
as well as new couplings are explored. It is also shown that these couplings can be verified at the level of EFT by either the combinations of Myers terms, pull-back, Taylor expanded of 
scalar fields or the mixed combination of the couplings of this paper as well as employed Bianchi identities.
For the first time, we also derive the algebraic and the complete form of the integrations  for some  arbitrary combinations of  Mandelstam variables and for the most general case 
$\int d^2z |1-z|^{a} |z|^{b} (z - \bar{z})^{c}
(z + \bar{z})^{3}$ on upper half plane as well.

 \end{quote}
\end{titlepage}

 \section{Introduction}

D-branes are  supposed to be  the fundamental objects that do exist in type II string theory.Indeed their roles have been 
enormously appreciating over the last decade, most notably by high energy physicist and in particular widely
by string theorists  \cite{Polchinski:1995mt,Witten:1995im}. 

\vskip.06in

To have some sort of  understanding
the brane's dynamics, we start off addressing various effective actions of these branes. Basically,  in the very 
stablished Dielectric effect  the issue of  multiple brane 's effective action as well as the appearance of 
commutator of two  massless transverse scalar fields (describing oscillations of branes)
were  clarified  \cite{Myers:1999ps}.  We have already applied the direct conformal field theory
methods and also made use of  the mixture of Ramond-Ramond (RR)-open string scattering 
amplitude computations to actually (up to some convincing field theory contents) 
gain the so called generalized Myers action.
\vskip.06in

It is worthwhile to point out  
this action  within its all order $\alpha'$ higher derivative corrections have been  investigated in \cite{Hatefi:2012zh}.
On the other hand,  already various  anomalous D-branes' couplings as well as dissolving branes inside 
the branes  have been verified in detail \cite{Green:1996dd}. Given the potential of S-matrix, various new 
couplings in  \cite{Hatefi:2015ora}  are  revealed  and the important point is that these couplings can not be 
established  by all the three standard ways of EFT, namely neither Taylor expansion, Myers terms nor pull-back 
formalism worked out.  Ultimately as  physical applications to those effective couplings , we first found out  the so called 
$N^3$ phenomenon for particularly $M5$ branes and some other systems , such as M2-M5 and consequently  dS solutions as well as 
   realizing the growth of the entropy of diverse configurations \cite{Hatefi:2012bp}.

\vskip.1in

The symmetrized action at non-Abelian level was given by \cite{Howe:2006rv}, where as originally the action for single bosonic brane had been introduced by Leigh in \cite{Leigh:1989jq}. Almost a decade later the supersymmetic part of the effective action  becomes known \cite{Cederwall:1996pv}.  To complete the effective actions 
various people including the author have taken a step further and verified within tremendous details the D-brane anti D-brane string theory effective actions that are consistent
with direct string amplitude calculations, while in this context duality does not seem to be promising any more, given the nature of tachyonic systems \cite{Michel:2014lva}.
\vskip.1in

Let us elaborate on the fact that the couplings of RR with even non-supersymmetric branes
at first glimpse investigated by the same prospective in  \cite{Kennedy:1999nn} as we head off from now
on. Further explanations about standard EFT couplings as well as almost all  effective actions can be  achieved
 in \cite{Hatefi:2012wj}.

\vskip.08in

The paper  is written  as follows. 
In the next section we try to find out all the closed form of the correlators for two transverse scalar fields, one gauge 
field and a potential RR, C-field $<V_{C^{-2}}  V_{A^{0}} V_{\phi^ {0}}  V_{\phi^ {0}} >$ we then continue
by just mentioning the final result of the same field contents but in symmetric picture.  
Although recently a method is given in \cite{Sen:2015hia}, we would like to evidently keep 
considering all the terms including all momenta of RR and in particular its momenta within the bulk such as 
$p.\xi_1$ and  $p.\xi_2$ terms inside the S-matrix.  Because  indeed various correlation functions such as  $<e^{ip.x(z)} \partial_i x^i(x_1)> $  obviously have non zero contributions to amplitude. We also clearly take into account the Bianchi identities to be able to produce new bulk singularities as well as all contact terms that are located in transverse directions.

For the first time, the explicit form of integrations on upper half plane for 
arbitrary combinations of Mandelstam variables and for generic case $\int d^2z |1-z|^{a} |z|^{b} (z - \bar{z})^{c}
(z + \bar{z})^{3}$ is achieved. We also  generate an infinite number of $t,s $ channel bulk singularity structures by means of all order $\alpha'$ corrections to pull-back of brane and highlight the fact that unlike  
\cite{Hatefi:2016enc}, neither there are $u$- channel bulk nor $(t+s+u)$- channel bulk singularity structures.
Due to presence of complete form of the S-matrix of this paper, several new contact interaction couplings will be discovered
and those couplings can just be verified at the level of Effective Field Theory (EFT) by either the combination of Myers terms, pull-back, Taylor expanded of  scalar fields or the mixed combination of the desired couplings  as we will point out later on.
Notice to the  extremely important point that by just carrying out direct S-matrix computations apart from exploring 
new couplings with distinguished structures, we are also able to precisely fix the coefficients of 
those new couplings to all orders in $\alpha'$.
 
 \section{ The  $<C^{-2} A^0  \phi^0 \phi^0>$ S-matrix}

First of all let us clarify the notation. $\mu,\nu = 0, 1,..., 9$ representing the whole space-time ,  world volume indices are 
$ b, c = 0, 1,..., p$ and eventually the transverse indices can be shown by  $i,j = p + 1,...,9$.
 In order to actually find out exact and all order $\alpha'$ contact terms as well as bulk singularity structures of BPS strings , one must apply direct CFT techniques to get to the complete form of the so called S-matrix elements. In this section we would like to  investigate the closed and complete form of  a particular BPS string amplitudes, a world volume gauge field and two transverse scalar fields in the presence of a potential $(p+1)$- form field of $C$- term  which is called potential of Ramond-Ramond (RR) in the whole space-time. This amplitude can be explored if one does all the CFT correlation functions of 
$<C^{-2} A^0  \phi^0 \phi^0>$ S-matrix. The interested reader  may find  some partial results that have come out of the precise 
and direct string scattering amplitude calculations in \cite{Barreiro:2013dpa}. Hence, one needs to entirely figure out all the correlators   $\lan V_{A}^{(0)}{(x_{1})}
V_{\phi}^{(0)}{(x_{2})}V_{\phi}^{(0)}{(x_{3})}
V_{RR}^{(-\frac{3}{2},-\frac{1}{2})}(z,\bar{z})\ran$ as well.\footnote{ Having regarded all RR momenta in bulk directions 
and the fact that winding modes are not covered in the whole ten dimensional flat space , one would get to know that definitely not all the elements of $<V_{C^{-2}} V_{A^0} V_{\phi^0}V_{\phi^0}>$ S-matrix can be explored from the recent  
 $<V_{C^{-2}} V_{\phi^0}V_{A^0} V_{A^0}>$ amplitude \cite{Hatefi:2016enc} , where the other explanations are given in  \cite{Hatefi:2012ve},\cite{Park:2008sg}.}


To do so, one simply separate all the bosonic and fermionic correlation functions and start to explore each of them. For discovering 
explicitly  all two spin operators  with different numbers of  fermion fields or currents, we also employ 
the  so called generalized  Wick-like rule, that is, the two point function of fermionic operators gets changed with a minus sign.\footnote{
$x_{ij}=x_i-x_j$, and $\alpha'=2$.}

Note that all the vertex operators, propagators  and on-shell relations are given in  section 2 of \cite{Hatefi:2016enc},
where the RR vertex operator in asymmetric picture was first hinted in \cite{Bianchi:1991eu} and eventually
its compact form is derived in \cite{Liu:2001qa}.
In order to simplify all the entire analysis of the correlation functions, we would like to write down just the compact form of 
the S-matrix as follows.
\beqa
{\cal A}^{C^{-2}A^0  \phi^0 \phi^0}&\sim&\Tr(\lam_1\lam_2\lam_3)\int dx_{1}dx_{2}
dx_{3}dx_{4}dx_{5}(P_{-}\fsC_{(n-1)}M_p)^{\al\be}I\xi_{1a}\xi_{2i}\xi_{3j}x_{45}^{-3/4}\nonumber\\&&\times
\bigg((x_{45}^{-5/4} C^{-1}_{\alpha\beta})\bigg[ a^a_1a^i_2 a^j_3 -\eta^{ij} x_{23}^{-2} a^a_1\bigg]+i\alpha'  k_{2b}
a^{a}_{1} a^{j}_{3} a^{ib}_{2}\nonumber\\&&
+i\alpha'  k_{1d} a^{ad}_{3} (-\eta^{ij} x_{23}^{-2}+a^i_2 a^j_3)-{\alpha'}^{2} k_{1d} k_{2b} a^j_3 a^{ibad}_{4}
+i\alpha'  k_{3c} a^a_1 a^i_2 a^{jc}_{5}\nonumber\\&&
-{\alpha'}^2 k_{3c} k_{2b} a^a_1 a^{jcib}_{6}-{\alpha'}^2 k_{3c} k_{1d} a^i_2  a^{jcad}_{7}-i{\alpha'}^3 k_{1d}k_{2b}k_{3c}
a_8^{jcibad}\bigg)
\labell{amp3q333},\eeqa
where
\beqa
I&=&|x_{12}|^{\alpha'^2 k_1.k_2}|x_{13}|^{\alpha'^2 k_1.k_3}|x_{14}x_{15}|^{\frac{\alpha'^2}{2} k_1.p}|x_{23}|^{\alpha'^2 k_2.k_3}|
x_{24}x_{25}|^{\frac{\alpha'^2}{2} k_2.p}
|x_{34}x_{35}|^{\frac{\alpha'^2}{2} k_3.p}|x_{45}|^{\frac{\alpha'^2}{4}p.D.p},\nonumber\\
a^i_2&=&ip^{i}\bigg(\frac{x_{54}}{x_{24}x_{25}}\bigg)\nonumber\\
a^a_1&=&ik_2^{a}\bigg(\frac{x_{42}}{x_{14}x_{12}}+\frac{x_{52}}{x_{15}x_{12}}\bigg)+ik_3^{a}
\bigg(\frac{x_{43}}{x_{14}x_{13}}+\frac{x_{53}}{x_{15}x_{13}}\bigg),\nonumber\\
a^j_3&=&ip^{j}\bigg(\frac{x_{54}}{x_{34}x_{35}}\bigg),\nonumber\\
a^{ib}_{2}&=&2^{-1}x_{45}^{-1/4}(x_{24}x_{25})^{-1} (\Gamma^{ib}C^{-1})_{\alpha\beta} ,\nonumber\\
a^{ad}_{3}&=&2^{-1}x_{45}^{-1/4}(x_{14}x_{15})^{-1}(\Gamma^{ad}C^{-1})_{\alpha\beta} ,\nonumber\\
a^{ibad}_{4}&=&2^{-2}x_{45}^{3/4}(x_{14}x_{15}x_{24}x_{25})^{-1}\bigg\{(\Gamma^{ibad}C^{-1})_{\alpha\beta}+
\alpha' p_1\frac{Re[x_{14}x_{25}]}{x_{12}x_{45}}\bigg\}
,\nonumber\\
a^{jc}_5&=&2^{-1}x_{45}^{-1/4}(x_{34}x_{35})^{-1} (\Gamma^{jc}C^{-1})_{\alpha\beta} ,\nonumber\\
a^{jcib}_6&=&2^{-2}x_{45}^{3/4}(x_{34}x_{35}x_{24}x_{25})^{-1}\bigg\{(\Gamma^{jcib}C^{-1})_{\alpha\beta}+
\alpha' p_2\frac{Re[x_{24}x_{35}]}{x_{23}x_{45}}+\alpha'^2 p_3\bigg(\frac{Re[x_{24}x_{35}]}{x_{23}x_{45}}\bigg)^{2}\bigg\},\nonumber\\
a^{jcad}_{7}&=&2^{-2}x_{45}^{3/4}(x_{34}x_{35}x_{14}x_{15})^{-1}\bigg\{(\Gamma^{jcad}C^{-1})_{\alpha\beta}
+\alpha' p_4\frac{Re[x_{14}x_{35}]}{x_{13}x_{45}}\bigg\}
\nonumber\eeqa
with the following expressions for the above correlators
\beqa
p_1&=&\bigg(\eta^{db}(\Gamma^{ia}C^{-1})_{\alpha\beta}
-\eta^{ab}(\Gamma^{id}C^{-1})_{\alpha\beta}\bigg),\nonumber\\
p_2&=&\bigg(\eta^{bc}(\Gamma^{ji}C^{-1})_{\alpha\beta}
+\eta^{ij}(\Gamma^{cb}C^{-1})_{\alpha\beta}\bigg),\nonumber\\
p_3&=&(C^{-1})_{\alpha\beta}\bigg(-\eta^{bc}\eta^{ij}\bigg),\nonumber\\
p_4&=&\bigg(\eta^{dc}(\Gamma^{ja}C^{-1})_{\alpha\beta}-\eta^{ac}(\Gamma^{jd}C^{-1})_{\alpha\beta}\bigg),\nonumber
\eeqa
The last fermionic correlation function  is  
\beqa
a_8^{jcibad}&=&<:S_{\al}(x_4):S_{\be}(x_5)::\psi^d\psi^a(x_1):\psi^b\psi^i(x_2):\psi^c\psi^j(x_3)>\label{33312}\eeqa
which does have  various terms and can be eventually found as follows
\beqa
a_8^{jcibad}&=&
\bigg\{(\Gamma^{jcibad}C^{-1})_{{\alpha\beta}}+\alpha' p_5\frac{Re[x_{14}x_{25}]}{x_{12}x_{45}}+
\alpha' p_6\frac{Re[x_{14}x_{35}]}{x_{13}x_{45}}+\alpha' p_7\frac{Re[x_{24}x_{35}]}{x_{23}x_{45}}
\nonumber\\&&+\alpha'^2 p_{8}\bigg(\frac{Re[x_{14}x_{25}]}{x_{12}x_{45}}\bigg)\bigg(\frac{Re[x_{14}x_{35}]}{x_{13}x_{45}}\bigg)
+\alpha'^2 p_9 \bigg(\frac{Re[x_{14}x_{25}]}{x_{12}x_{45}}\bigg)\bigg(\frac{Re[x_{24}x_{35}]}{x_{23}x_{45}}\bigg)
\nonumber\\&&+\alpha'^2 p_{10} \bigg(\frac{Re[x_{14}x_{35}]}{x_{13}x_{45}}\bigg)\bigg(\frac{Re[x_{24}x_{35}]}{x_{23}x_{45}}\bigg)
+\alpha'^2 p_{11}\bigg(\frac{Re[x_{24}x_{35}]}{x_{23}x_{45}}\bigg)^{2}
\nonumber\\&&
+\alpha'^3 p_{12} \bigg(\frac{Re[x_{24}x_{35}]}{x_{23}x_{45}}\bigg)\bigg(\frac{Re[x_{14}x_{25}]}{x_{12}x_{45}}\bigg)
\bigg(\frac{Re[x_{14}x_{35}]}{x_{13}x_{45}}\bigg)
\bigg\}
\label{hh333}2^{-3}x_{45}^{7/4}(x_{14}x_{15}x_{24}x_{25}x_{34}x_{35})^{-1}\nonumber
\eeqa
where
\beqa
p_5&=&\bigg(\eta^{db}(\Gamma^{jcia}C^{-1})_{\alpha\beta}
-\eta^{ab}(\Gamma^{jcid}C^{-1})_{\alpha\beta}\bigg),\nonumber\\
p_6&=&\bigg(\eta^{dc}(\Gamma^{jiba}C^{-1})_{\alpha\beta}
-\eta^{ac}(\Gamma^{jibd}C^{-1})_{\alpha\beta}\bigg),\nonumber\\
p_7&=&\bigg(\eta^{bc}(\Gamma^{jiad}C^{-1})_{\alpha\beta}
+\eta^{ij}(\Gamma^{cbad}C^{-1})_{\alpha\beta}\bigg),\nonumber\\
p_8&=&\bigg(\eta^{db}\eta^{ac}(\Gamma^{ji}C^{-1})_{\alpha\beta}
-\eta^{dc}\eta^{ab}(\Gamma^{ji}C^{-1})_{\alpha\beta}\bigg),\nonumber\\
p_9&=&\bigg(\eta^{db}\eta^{ij}(\Gamma^{ca}C^{-1})_{\alpha\beta}-\eta^{ab}\eta^{ij}(\Gamma^{cd}C^{-1})_{\alpha\beta}\bigg),\nonumber\\
p_{10}&=&\bigg(-\eta^{dc}\eta^{ij}(\Gamma^{ba}C^{-1})_{\alpha\beta}+\eta^{ac}\eta^{ij}(\Gamma^{bd}C^{-1})_{\alpha\beta}\bigg)
\nonumber\\
p_{11}&=&\bigg(-\eta^{bc}\eta^{ij}(\Gamma^{ad}C^{-1})_{\alpha\beta}\bigg),\nonumber\\
p_{12}&=&(C^{-1})_{\alpha\beta}\bigg(\eta^{ij}\eta^{ab}\eta^{dc}-\eta^{ij}\eta^{ac}\eta^{db}\bigg),
\eeqa

Note that we wrote the amplitude in a manifest way so that, one is able to explicitly check that the amplitude is invariant
under SL(2,R) transformation and  volume of conformal killing group can be cancelled off  by fixing three positions of
space-time, where we choose just to fix all the locations of open strings.\footnote{ $  x_{1}=0 ,x_{2}=1,x_{3}\rightarrow \infty$ .}
Thus , one needs to integrate out the remaining moduli space which is upper half plane and indeed all the integrations related to
RR location \cite{Fotopoulos:2001pt}. Note that all the details of integrations are explained in Appendix B of \cite{Hatefi:2012wj}. 
However, in order to explore the integrations for $p_3$  of $a^{jcib}_{6}$ and $p_{12}$  of $a^{jcibad}_{8}$ , 
one must find the algebraic solution of the following integrals $ 
 \int d^2 \!z |1-z|^{a} |z|^{b} (z - \bar{z})^{c} (z + \bar{z})^{3}$
so that all  $a,b,c$ are  written down in terms of any arbitrary Mandelstam variables. Once we are dealing with 
$(z + \bar{z})$  the result  is explored in
\cite{Fotopoulos:2001pt}, meanwhile for $(z + \bar{z})^{2}$  one derives the  entire result from \cite{Hatefi:2012wj}. \footnote{
where definitions are \beqa
s&=&\frac{-\alpha'}{2}(k_1+k_3)^2,\quad t=\frac{-\alpha'}{2}(k_1+k_2)^2,\quad u=\frac{-\alpha'}{2}(k_2+k_3)^2
\nonumber\eeqa.} For the first time, one finds the algebraic solution for the above integrals for $d=3$ as follows:
\beqa
 \int d^2 \!z |1-z|^{a} |z|^{b} (z - \bar{z})^{c}
(z + \bar{z})^{3}&=& (2i)^c 2^3 \pi \frac{K_1+K_2}{\Gamma(\frac{-a}{2})\Gamma(\frac{-b}{2})\Gamma(\frac{(a+b)}{2}+c+5)}
\eeqa

where the functions
 $K_1,K_2$ are
\beqa
K_1&=& \Gamma(1+\frac{(a+c)}{2})\Gamma(4+\frac{(b+c)}{2})\Gamma(-1-\frac{(a+b+c)}{2})\Gamma(\frac{1+c}{2}),\nonumber\\
K_2&=& \frac{3}{2}\Gamma(2+\frac{(a+c)}{2})\Gamma(3+\frac{(b+c)}{2})\Gamma(-2-\frac{(a+b+c)}{2})\Gamma(\frac{1+c}{2}),
\label{L1s}
\eeqa

Having set the solution for the new integrals , one would be able to  obtain the final result for the S-matrix element in an asymmetric picture  as follows

\beqa {\cal A}^{C^{-2} A^{0}  \phi^{0} \phi^{0} }&=&{\cal A}_{1}+{\cal A}_{2}+{\cal A}_{3}+{\cal A}_{41}+{\cal A}_{42}
+{\cal A}_{5}\nonumber\\&&{\cal A}_{61}+{\cal A}_{62}
+{\cal A}_{63}+{\cal A}_{64}
+{\cal A}_{71}+{\cal A}_{72}+{\cal A}_{81}+{\cal A}_{82}\nonumber\\&&{\cal A}_{83}+{\cal A}_{84}
+{\cal A}_{85}+{\cal A}_{86}+{\cal A}_{87}+{\cal A}_{88}+{\cal A}_{89}
\labell{711u}\eeqa
where
\beqa
{\cal A}_{1}&\!\!\!\sim\!\!\!&i  \Tr(P_{-}\fsC_{(n-1)}M_p) 
\bigg[-2suk_2.\xi_1 p.\xi_2 p.\xi_3L_1 +2tu k_3.\xi_1 p.\xi_2 p.\xi_3L_1\nonumber\\&&
+2sk_2.\xi_1 \xi_3.\xi_2 L_2(-s-u)(-t-u)
-2tk_3.\xi_1 \xi_3.\xi_2 L_2(-s-u)(-t-u)\bigg],
\nonumber\\
{\cal A}_{2}&\sim&i k_{2b}\xi_{2i}p.\xi_3\Tr(P_{-}\fsC_{(n-1)}M_p \Gamma^{ib}) L_1
\bigg\{2us k_2.\xi_{1} -2ut k_3.\xi_{1}\bigg\}\nonumber\\
{\cal A}_{3}&\sim&i k_{1d}\xi_{1a}\Tr(P_{-}\fsC_{(n-1)}M_p \Gamma^{ad}) 
\bigg[-L_3\frac{(-s-u)(-t-u)}{(-u-\frac{1}{2}) } \xi_{3}.\xi_{2} -p.\xi_2 p.\xi_3L_4\bigg]\nonumber\\
{\cal A}_{41}&\sim&i p.\xi_3 \Tr(P_{-}\fsC_{(n-1)}M_p \Gamma^{ibad})\xi_{2i}\xi_{1a} k_{2b}k_{1d} L_4 
\nonumber\\
{\cal A}_{42}&\sim& i p.\xi_3   \xi_{2i} \Tr(P_{-}\fsC_{(n-1)}M_p \Gamma^{ia}) L_1\bigg\{tsu \xi_{1a}+2su k_2.\xi_{1} k_{1a}
\bigg\}
\nonumber\\
{\cal A}_{5}&\sim&i p.\xi_2 k_{3c}\xi_{3j}\Tr(P_{-}\fsC_{(n-1)}M_p \Gamma^{jc}) L_1
\bigg\{2 su k_2.\xi_{1}-2tu k_3.\xi_1\bigg\}\nonumber\\
{\cal A}_{61}&\sim&i \Tr(P_{-}\fsC_{(n-1)}M_p \Gamma^{jcib})\xi_{2i}\xi_{3j} k_{3c}k_{2b}L_1 (-2suk_2.\xi_1+2tuk_3.\xi_1)\nonumber\\
{\cal A}_{62}&\sim& iL_3\Tr(P_{-}\fsC_{(n-1)}M_p \Gamma^{ji})\xi_{2i} \xi_{3j}  \bigg\{2suk_2.\xi_1-2tuk_3.\xi_1\bigg\}
\nonumber\\
{\cal A}_{63}&\sim& iL_3\Tr(P_{-}\fsC_{(n-1)}M_p \Gamma^{cb})\xi_{2}.\xi_{3} (-k_{2b}k_{3c}) \bigg\{4sk_2.\xi_1-4tk_3.\xi_1\bigg\}
\nonumber\\
{\cal A}_{64}&\sim& -iu (2st-u)\xi_{2}.\xi_{3} \Tr(P_{-}\fsC_{(n-1)}M_p )L_2  (-2sk_2.\xi_1+2tk_3.\xi_1) 
\nonumber\\
{\cal A}_{71}&\sim&i p.\xi_2 \Tr(P_{-}\fsC_{(n-1)}M_p \Gamma^{jcad})\xi_{1a}\xi_{3j}  k_{3c}k_{1d} L_4\nonumber\\
{\cal A}_{72}&\sim& i p.\xi_2 \Tr(P_{-}\fsC_{(n-1)}M_p \Gamma^{ja}) L_1 \xi_{3j}\bigg\{-uts\xi_{1a} 
-2ut k_3.\xi_1 k_{1a}\bigg\}
\nonumber\\
{\cal A}_{81}&\sim& -i \Tr(P_{-}\fsC_{(n-1)}M_p \Gamma^{jcibad})\xi_{2i}\xi_{1a}\xi_{3j}k_{1d} k_{2b}k_{3c} L_4\nonumber\\
{\cal A}_{82}&\sim& -i su\Tr(P_{-}\fsC_{(n-1)}M_p \Gamma^{jcia})\xi_{2i}\xi_{3j} k_{3c} L_1 \bigg\{t \xi_{1a}+2k_2.\xi_{1} k_{1a}
\bigg\}
\nonumber\\
{\cal A}_{83}&\sim&  -i tu\Tr(P_{-}\fsC_{(n-1)}M_p \Gamma^{jiba})\xi_{3j}\xi_{2i}k_{2b} L_1\bigg\{-s \xi_{1a}-2k_3.\xi_{1}
k_{1a}\bigg\}
\nonumber\\
{\cal A}_{84}&\sim& -ist  k_{1d}\xi_{1a} L_1\bigg\{u\xi_{2i}\xi_{3j}\Tr(P_{-}\fsC_{(n-1)}M_p \Gamma^{jiad})
-2\xi_{2}.\xi_{3}k_{3c} k_{2b}\Tr(P_{-}\fsC_{(n-1)}M_p \Gamma^{cbad})\bigg\}
\nonumber\\
{\cal A}_{85}&\sim&-i \Tr(P_{-}\fsC_{(n-1)}M_p \Gamma^{ji}) \xi_{2i} \xi_{3j} L_3\bigg\{-2tu k_{3}.\xi_{1}+2su k_2.\xi_{1}\bigg\}
\nonumber\\\nonumber\\
{\cal A}_{86}&\sim& -i\xi_{3}.\xi_{2} \Tr(P_{-}\fsC_{(n-1)}M_p \Gamma^{ca})L_3 \bigg\{2ts k_{3c}  \xi_{1a}+4sk_2.\xi_1 k_{3c}k_{1a}\bigg\}\nonumber\\
{\cal A}_{87}&\sim& -i\xi_{3}.\xi_{2} \Tr(P_{-}\fsC_{(n-1)}M_p \Gamma^{ba})L_3 \bigg\{2ts k_{2b}  \xi_{1a}+4tk_3.\xi_1 k_{2b}
k_{1a}\bigg\}\nonumber\\
{\cal A}_{88}&\sim& i L_3 \Tr(P_{-}\fsC_{(n-1)}M_p \Gamma^{ad})\xi_{1a} k_{1d} (u\xi_2.\xi_3)  \bigg\{\frac{-2st+u+s+t}{(-u-\frac{1}{2})}\bigg\}
\nonumber\\
{\cal A}_{89}&\sim&-i\xi_2.\xi_3 L_2 \Tr(P_{-}\fsC_{(n-1)}M_p) 
 (2s k_2.\xi_1-2t k_3.\xi_1)\bigg(tu+s(t+u+2tu)\bigg)
\labell{483333}\eeqa
with the definition for  the functions
 $L_1,L_2,L_3,L_4$ as follows
\beqa
L_1&=&(2)^{-2(t+s+u)}\pi{\frac{\Gamma(-u)
\Gamma(-s)\Gamma(-t)\Gamma(-t-s-u+\frac{1}{2})}
{\Gamma(-u-t+1)\Gamma(-t-s+1)\Gamma(-s-u+1)}}\nonumber\\
L_2&=&(2)^{-2(t+s+u+1)}\pi{\frac{\Gamma(-u)
\Gamma(-s)\Gamma(-t)\Gamma(-t-s-u-\frac{1}{2})}
{\Gamma(-u-t+1)\Gamma(-t-s+1)\Gamma(-s-u+1)}}\nonumber\\
L_3&=&(2)^{-2(t+s+u)-1}\pi{\frac{\Gamma(-u+\frac{1}{2})
\Gamma(-s+\frac{1}{2})\Gamma(-t+\frac{1}{2})\Gamma(-t-s-u)}
{\Gamma(-u-t+1)\Gamma(-t-s+1)\Gamma(-s-u+1)}},\nonumber\\
L_4&=&(2)^{-2(t+s+u)+1}\pi{\frac{\Gamma(-u+\frac{1}{2})
\Gamma(-s+\frac{1}{2})\Gamma(-t+\frac{1}{2})\Gamma(-t-s-u+1)}
{\Gamma(-u-t+1)\Gamma(-t-s+1)\Gamma(-s-u+1)}},\nonumber\\
\label{Ls89}
\eeqa

 Lets elaborate on the details. In fact the sum of 
 the 3rd term of ${\cal A}_{1}$, 1st term of ${\cal A}_{64}$ and the 1st term of ${\cal A}_{89}$  is zero as well as 
 the sum of  the 4th term of ${\cal A}_{1}$, 2nd term of ${\cal A}_{64}$ and the 2nd term of ${\cal A}_{89}$ . This obviously means that
 the last two terms of ${\cal A}_{1}$, the entire  ${\cal A}_{64}$  and the whole ${\cal A}_{89}$ have no contribution
 to the asymmetric S-matrix at all.
 
 Note that ${\cal A}_{62}$ is precisely cancelled off with the entire terms inside ${\cal A}_{85}$ , this also clarifies that
${\cal A}_{62},{\cal A}_{85}$ will not contribute to our physical asymmetric amplitude either.
On the other hand , if we just consider the RR in terms of its field strength we get to obtain the following \cite{Hatefi:2012zh}
\beqa {\cal A}^{<C^{-1}A^{-1}\phi^{0}\phi^{0}>}&=&{\cal A}_{1}+{\cal A}_{2}+{\cal A}_{3}+{\cal A}_{4}+{\cal A}_{5}+{\cal A}_{6}
+{\cal A}_{7}+{\cal A}_{8}+{\cal A}_{9}+{\cal A}_{10}\labell{11u89}\eeqa
where
\beqa
{\cal A}_{1}&\!\!\!\sim\!\!\!&-2^{-1/2}\xi_{1a}\xi_{2i}\xi_{3j}
\bigg[k_{3c}k_{2b}\Tr(P_{-}\fsH_{(n)}M_p\Gamma^{jciba})-k_{2b}p^j\Tr(P_{-}\fsH_{(n)}M_p\Gamma^{iba})\nonumber\\&&-k_{3c}p^i\Tr(P_{-}\fsH_{(n)}M_p\Gamma^{jca})+p^ip^j\Tr(P_{-}\fsH_{(n)}M_p\gamma^{a})\bigg]
4(-s-t-u)L_3,
\nonumber\\
{\cal A}_{2}&\sim&2^{-1/2}
\bigg\{-
2\xi_{1}.k_{2}k_{3c}\xi_{3j}\xi_{2i}\Tr(P_{-}\fsH_{(n)}M_p \Gamma^{jci})\bigg\}(us)L_1\nonumber\\
{\cal A}_{3}&\sim&2^{-1/2}
\bigg\{\xi_{1a}\xi_{2i}\xi_{3j}\Tr(P_{-}\fsH_{(n)}M_p \Gamma^{jia})\bigg\}(-ust)L_1\nonumber\\
{\cal A}_{4}&\sim&2^{-1/2}
\bigg\{
2k_{3}.\xi_{1}k_{2b}\xi_{3j}\xi_{2i}\Tr(P_{-}\fsH_{(n)}M_p \Gamma^{jib})\bigg\}(ut)L_1\nonumber\\
\nonumber\\
{\cal A}_{5}&\sim&2^{-1/2}
\bigg\{2\xi_{3}.\xi_{2}k_{2b}k_{3c}\xi_{1a}\Tr(P_{-}\fsH_{(n)}M_p \Gamma^{cba})\bigg\}(st) L_1\nonumber\\
{\cal A}_{6}&\sim& 2^{1/2}(us) L_{1}\bigg\{p^j\xi_1.k_2\xi_{2i}\xi_{3j}\Tr(P_{-}\fsH_{(n)}M_p\gamma^i)
\bigg\}
\nonumber\\
{\cal A}_{7}&\sim&-2^{-1/2} (ut) L_1\bigg\{
2k_{3}.\xi_1p^i\xi_{3j}\xi_{2i}\Tr(P_{-}\fsH_{(n)}M_p\gamma^j)\bigg\}
\nonumber\\
{\cal A}_{8}&\sim&2^{1/2}L_3\bigg\{2k_2.\xi_1 k_{3c}\Tr(P_{-}\fsH_{(n)}M_p\gamma^c)
(-s\xi_2.\xi_3)\bigg\}.
\nonumber\\
{\cal A}_{9}&\sim&2^{1/2}L_3\bigg\{2k_3.\xi_{1}k_{2b}\Tr(P_{-}\fsH_{(n)}M_p\gamma^b)
(-t\xi_2.\xi_3)\bigg\}
\nonumber\\
{\cal A}_{10}&\sim&2^{1/2}L_3\bigg\{\xi_{1a}\Tr(P_{-}\fsH_{(n)}M_p\gamma^a)
(ts\xi_3.\xi_2)\bigg\}
\labell{48089}\eeqa
While the other symmetric amplitude has already been found in \cite{Hatefi:2015ora}  to be
\beqa {\cal A}^{<C ^{-1}A^{0}\phi ^{-1}\phi^{0}>}&=&{\cal A}_{1}+{\cal A}_{2}+{\cal A}_{3}+{\cal A}_{4}+{\cal A}_{5}
+{\cal A}_{6}
\labell{7111289u}\eeqa
where
\beqa
{\cal A}_{1}&\!\!\!\sim\!\!\!&2^{-1/2}\xi_{1a}\xi_{2i}\xi_{3j} p^j\Tr(P_{-}\fsH_{(n)}M_p\gamma^{i})
\bigg[-2k^a_3(ut)+2k^a_2(us)\bigg]L_1
\nonumber\\
{\cal A}_{2}&\sim&2^{-1/2}k_{3c}\bigg\{-2k_2.\xi_1 \xi_{2i}\xi_{3j}(us)L_1\Tr(P_{-}\fsH_{(n)}M_p \Gamma^{jci})
+2k_3.\xi_1 \xi_{2i}\xi_{3j}(ut)L_1\Tr(P_{-}\fsH_{(n)}M_p \Gamma^{jci})
\nonumber\\&&+4t\xi_2.\xi_3 k_3.\xi_1L_3\Tr(P_{-}\fsH_{(n)}M_p \gamma^{c})
-4s\xi_2.\xi_3 k_2.\xi_1L_3\Tr(P_{-}\fsH_{(n)}M_p \gamma^{c})\bigg\}\nonumber\\
{\cal A}_{3}&\sim&2^{-1/2} k_{1b}\xi_{1a}\xi_{2i}\xi_{3j}4(-u-s-t)L_3\bigg(\Tr(P_{-}\fsH_{(n)}M_p\Gamma^{iab}) p^j-k_{3c}\Tr(P_{-}\fsH_{(n)}M_p\Gamma^{jciab})\bigg)\nonumber\\
{\cal A}_{4}&\sim&2^{-1/2}(ut)L_1
\bigg\{ -s\xi_{1a}\xi_{2i}\xi_{3j}\Tr(P_{-}\fsH_{(n)}M_p \Gamma^{jia})-2k_3.\xi_1 k_{1b}\xi_{2i}\xi_{3j}\Tr(P_{-}\fsH_{(n)}M_p \Gamma^{jib})\bigg\}\nonumber\\
{\cal A}_{5}&\sim&2^{1/2}(st)L_1\xi_2.\xi_3\xi_{1a}k_{1b}k_{3c}\Tr(P_{-}\fsH_{(n)}M_p \Gamma^{cab})\nonumber\\
{\cal A}_{6}&\sim&2^{1/2}\xi_{3}.\xi_{2}\bigg(ts\Tr(P_{-}\fsH_{(n)}M_p \gamma^{a})\xi_{1a}+2tk_3.\xi_1
\Tr(P_{-}\fsH_{(n)}M_p \gamma^{b})k_{1b}\bigg)L_3
\labell{48376589}\eeqa
where the functions
 $L_1,L_2$ are given in \reef{Ls89}. In the next section we are going to compare within details all 
 the singularity structures of asymmetric with symmetric analysis and then start producing an
 infinite number of $t,s$-channel bulk singularity structures in an EFT as well.
\section{ Singularity Comparisons }

In this section we are going to provide precise analysis of all singularity structures involving  
even bulk singularities that are about to be found in this paper.
To do so, we first try to regenerate singularities that have been already derived in symmetric analysis.

In order to produce all infinite t-channel poles of symmetric result, one needs to  start adding up the first term  of 
${\cal A}_{61}$ with the second term of ${\cal A}_{82}$ and apply momentum conservation along the brane to obtain 
\beqa
2isu k_2.\xi_1 L_1 \Tr(P_{-}\fsC_{(n-1)}M_p \Gamma^{jcid})\xi_{2i}\xi_{3j} k_{3c} (p+k_3)_{d}  \eeqa
obviously the 2nd term in above equation has no contribution to S-matrix , because it is symmetric under interchanging
$k_{3c}, k_{3d}$ but also is antisymmetric as it involves $\epsilon $ tensor so the result for the 2nd term is zero, meanwhile
 its first term $(p\fsC=\fsH)$ does generate ${\cal A}_{2}$ of \reef{11u89} ( which is the fifth term  of S-matrix elements in 
 symmetric picture).
One can do the same procedure , namely by adding the 2nd terms of ${\cal A}_{83},{\cal A}_{61}$ and using momentum
conservation, we gain all infinite s-channel poles or 
${\cal A}_{4}$ of \reef{11u89} as follows 
\beqa
-2itu k_3.\xi_1 L_1 \Tr(P_{-}\fsC_{(n-1)}M_p \Gamma^{jibd})\xi_{2i}\xi_{3j} k_{2b} (p+k_2)_{d}  \eeqa

Note that, making use of momentum conservation and $(p\fsC=\fsH)$, one reveals that
the 2nd term of ${\cal A}_{84}$ exactly constructs all infinite u-channel poles or ${\cal A}_{5}$ of \reef{11u89} as well. 

  \vskip.1in

Indeed  for  this particular 
$<C^{-2}A^{0}\phi^{0}\phi^{0}>$ S-matrix, we have evidently shown that there are no  $u$-channel Bulk singularity structures at all. 
The physical explanation for this is as follows.  Suppose, we take into account the following rule in effective field theory side ,

 \beqa
{\cal A}&=&V^a_{\alpha}(C_{p-3},A_1,A)G^{ab}_{\alpha\beta}(A)V^b_{\beta}(A,\phi_2,\phi_3),\label{amp644}
\eeqa

we then may clarify that the vertex of $V^a_{\alpha}(C_{p-3},A_1,A)$ must be derived from Chern-Simons
coupling as $(2\pi\alpha')^2\int_{\Sigma_{p+1}} C_{p-3}\wedge F\wedge F$ and in fact all $(p+1)$  indices have 
been considered and there are no left over indices to be compensated by transverse directions
(we have no external scalar field for this part of the sub field theory amplitude), which is why we no longer have any $u$-channel bulk singularity structures. 

\vskip.1in

All u-channel gauge field poles can be written as 
\beqa
\mu_p(2\pi\alpha')^{2} 2k_{2b} k_{3c} p_{d}\xi_2.\xi_3\frac{1}{(p-3)!u}\eps^{a_{0}\cdots a_{p-4}cbad} C_{a_{0}\cdots a_{p-4}}\xi_{1a}\sum_{n=-1}^{\infty}b_n\bigg(\frac{\alpha'}{2}\bigg)^{n+1}(s+t)^{n+1}
\label{ope23}\eeqa

Considering \reef{amp644}, taking the fixed scalar fields's kinetic term in the action  (it receives no  correction at all ) as  
  $T_p \frac{(2\pi\alpha')^2}{2}\Tr(D^a\phi^iD_a\phi_i)$   , one finds out  the 
$V_{\beta}^{b}(A,\phi_2,\phi_3)$  and gauge field propagator.\footnote{
\beqa
V_{\beta}^{b}(A,\phi_2,\phi_3)&=&i(2\pi\alpha')^2 T_p
  \xi_2.\xi_3 (k_2-k_3)^b \Tr(\lambda_2\lambda_3\lambda_\beta)
\nonumber\\
G_{\alpha\beta}^{ab}(A)&=&\frac{-i}{(2\pi\alpha')^2 T_p}\frac{\delta^{ab}
\delta_{\alpha\beta}}{k^2}\,\,\, ,
\label{ver137}\eeqa}
By taking the higher derivative corrections to Chern-Simons coupling  as follows 

 \beqa
  i(2\pi\alpha')^2\mu_p\int d^{p+1}\sigma   \quad \sum_{n=-1}^{\infty}b_n (\alpha')^{n+1}\quad C_{p-3}\wedge D_{a_0}\cdots D_{a_n} F
  \wedge D^{a_0}\cdots D^{a_n} F
\eeqa
we would be able to exactly generate  the extension of the  $V^a_{\alpha}(C_{p-3},A_1,A)$  vertex operator to all order $\alpha'$ as below 
 \beqa
V^a_{\alpha}(C_{p-3},A_1,A)&=&\frac{(2\pi\alpha')^2\mu_p}{(p-3)!} \eps^{a_0\cdots a_{p-1}a} C_{a_0\cdots a_{p-4}}\xi_{1a_{p-3}}k_{a_{p-2}} p_{a_{p-1}}
\sum_{n=-1}^{\infty}b_n(t+s)^{n+1}
\label{990}\eeqa

where $k=(k_2+k_3)$  is employed and it now becomes clear that if we substitute \reef{990} and \reef{ver137} into field theory amplitude , then all  order u-channel singularities of  string amplitude in \reef{ope23} can be  explored in EFT as well.

\vskip.1in

Adding the 1st term of ${\cal A}_{63}$ with the 2nd term of ${\cal A}_{86}$ and applying momentum conservation, one explores
 \beqa
 4is k_2.\xi_1 \xi_{3}.\xi_{2} L_3 \Tr(P_{-}\fsC_{(n-1)}M_p \Gamma^{cd}) k_{3c}  (p+k_3)_{d} \eeqa
 where the second term in above equation  has no contribution to S-matrix, while its first term does generate precisely
 ${\cal A}_{8}$ of symmetric result in \reef{11u89}  ( of course with a different sign, which is over all factor at the end ).

Having added up  the 2nd terms of  ${\cal A}_{87},{\cal A}_{63}$, we were able to obtain the following term 
   \beqa
 4it k_3.\xi_1 \xi_{3}.\xi_{2} L_3 \Tr(P_{-}\fsC_{(n-1)}M_p \Gamma^{bd}) k_{2b}  (p+k_2)_{d} \eeqa
which is exactly ${\cal A}_{9}$ of symmetric result of \reef{11u89} (with a different sign).

Eventually if we add up all the first terms of ${\cal A}_{3},{\cal A}_{86},{\cal A}_{87}$ with the entire ${\cal A}_{88}$, we derive 
 \beqa
 -2its \xi_{1a} \xi_{3}.\xi_{2} L_3 \Tr(P_{-}\fsC_{(n-1)}M_p \Gamma^{da})   (k_1+k_2+k_3)_{d} \eeqa
  Now using momentum conservation and $(p\fsC=\fsH)$, one is able to regenerate precisely ${\cal A}_{10}$ of symmetric result 
  of \reef{11u89}. Thus all the 
  infinite $(t+s+u)$ channel poles have also been reconstructed. Note to the following important point.
  
  \vskip.1in

Indeed  here for  this particular 
$<C^{-2}A^{0}\phi^{0}\phi^{0}>$ S-matrix  (unlike $<C^{-2}\phi^{0}A^{0}A^{0}>$  S-matrix ), we have clearly shown that there 
is no even one $(t+s+u)$-channel Bulk singularity structure. The physical explanation for that is as follows. 
Suppose, we consider the following rule in effective field theory side ,
 \beqa
{\cal A}&=&V_{\alpha}^{a}(C_{p-1},A)G_{\alpha\beta}^{ab}(A)V_{\beta}^{b}(A,A_1,\phi_2,\phi_3)\nonumber\eeqa

then one observes that the vertex of $V_{\alpha}^{a}(C_{p-1},A)$ must be derived from Chern-Simons coupling as 
$(2\pi\alpha')\int_{\Sigma_{p+1}} C_{p-1}\wedge F$ and in fact all $(p+1)$  indices have been taken into account and
there are no left over indices to be compensated by transverse directions either,   that is why we  have no $(t+s+u)$-channel
bulk singularity structures any more.  It is worth noting that the universal conjecture of all order $\alpha'$ corrections in \cite{Hatefi:2012rx} has played the significant role in matching all Supersymmetric Yang-Mills couplings at both string and EFT levels.

\vskip.1in

Now if we consider two gauge field two scalar couplings to   all order in $\alpha'$
(appeared in  \cite{Hatefi:2012ve})  and construct $V_{\beta}^{b}(A,A_1,\phi_2,\phi_3)$,
then we will be able to precisely generate all infinite gauge field of $(t+s+u)$ channels. 
These poles have already been constructed out in \cite{Hatefi:2012zh}, 
where we advise the reader to explore them directly in section four of \cite{Hatefi:2012zh}. 
Furthermore, for the same reasons , one immediately expects not to have  $u$- channel Bulk singularities either.

\vskip.1in

  Considering the 1st term of ${\cal A}_{2}$ of asymmetric amplitude and  the 2nd term of ${\cal A}_{42}$, 
  applying momentum conservation and taking $(p\fsC=\fsH)$ , not only we obtain the  
${\cal A}_{6}$ of symmetric result in  \reef{11u89}  
\beqa
-2isu L_1 k_2.\xi_{1} p.\xi_3   \xi_{2i} \Tr(P_{-}\fsC_{(n-1)}M_p \Gamma^{id})   (p+k_3)_{d} \label{p90}\eeqa
but also we generate a new kind of bulk pole. Indeed the 2nd term in \reef{p90} is related to an infinite number of 
t- channel  extra bulk poles, which will be taken care of.

Finally, by adding the 2nd terms of ${\cal A}_{5} $ and $ {\cal A}_{72}$ and making use of momentum conservation,
we produce the following terms 
\beqa
 2iut k_3.\xi_1 p.\xi_2  L_1 \xi_{3j}\Tr(P_{-}\fsC_{(n-1)}M_p \Gamma^{jc}) (p+k_2)_{c} \label{p93}\eeqa

where the first term in  \reef{p93} does produce ${\cal A}_{7}$ of symmetric result in 
\reef{11u89}, while the 2nd term in \reef{p93} is exactly an infinite number of 
s- channel  extra bulk poles for which  remain to be explored.
Notice that the first terms of ${\cal A}_{5}, {\cal A}_{1}$, also the second terms of ${\cal A}_{2}, {\cal A}_{1}$
 of asymmetric S-matrix  \reef{711u} do generate an infinite number of bulk $t,s$-channel singularities accordingly,
 where we consider them in the next sections as well.

\section{ All order t,s -channel Bulk Singularity structures }

As we have explicitly shown in the previous section,  we could precisely  produce all the singularities of 
\reef{11u89} by using some  (but definitely not all) of the singularities of
asymmetric S-matrix.  Indeed unlike the previous section, here not all the indices of Wess-Zumino action
can be covered by world volume indices and in fact due to presence of external scalar field
states  as well as all non zero $p.\xi_1,p.\xi_2$ terms, one expects to have bulk t,s channel singularities as well, for which
we discuss from now on.

\vskip.1in

All infinite massless scalar t,s channel singularities ( not Bulk  $t,s$-channel singularities ) have already been generated in section 4.1 of  \cite{Hatefi:2012zh} but  the aim of this section is to find out all order t,s channel Bulk singularity structures.

Performing  careful comparisons  of all singularities in both symmetric and antisymmetric amplitudes,
 as well as  extracting all the related traces, one would be able to write down 
 all order t-channel Bulk singularity structures  that do exist just in asymmetric S-matrix
 \reef{711u} as follows:
 
 \beqa
 2ius k_2.\xi_{1}\xi_{2i}\xi_{3j}  \frac{16 L_1}{(p+1)!}   \bigg\{  \epsilon^{a_{0}...a_{p}}
 \bigg(-p^i p^j   C_{ a_{0}...a_{p}}\bigg)\nonumber\\
+  k_{3c}
 \epsilon^{a_{0}...a_{p-1}c} \bigg(p^i C_{j a_{0}...a_{p-1}}-p^j C_{i a_{0}...a_{p-1}}\bigg)\bigg\} \label{lop11}\eeqa

 and also all order s- channel bulk singularities  as follows 
\beqa
 2iut k_3.\xi_{1}\xi_{2i}\xi_{3j}  \frac{16 L_1}{(p+1)!}   \bigg\{  \epsilon^{a_{0}...a_{p}}\bigg(p^i p^j   C_{ a_{0}...a_{p}}\bigg)\nonumber\\
+  k_{2b}
 \epsilon^{a_{0}...a_{p-1}b} \bigg(p^i C_{j a_{0}...a_{p-1}}-p^j C_{i a_{0}...a_{p-1}}\bigg)\bigg\} \label{lop22}\eeqa

Note that all infinite t,s channel bulk singularities of
\reef{lop11} and \reef{lop22}, are needed as we are going to produce them in an EFT by introducing various couplings as follows.



\vskip.1in

Here we just produce all the infinite t-channel bulk singularities of  \reef{lop11} 
and then according to symmetries and by exchanging the scalar fields's momenta $k_2 \leftrightarrow k_3$ and interchanging the scalar fields 
polarizations $\xi_2 \leftrightarrow  \xi_3$
one also will be able to explore all the infinite s-channel bulk singularities in an EFT as well.

\vskip.1in

Let us apply  $us L_1$ expansion to    \reef{lop11} to generate all infinite t- channel  Bulk singularity singularities as  follows 

 \beqa
 2i k_2.\xi_{1}    \frac{16 \pi^2 \mu_p}{(p+1)!} \sum_{n=-1}^{\infty}b_n\frac{1}{t}(u+s)^{n+1} \Tr(\lambda_1\lambda_2\lambda_3) \nonumber\times
\bigg\{  \bigg(- \epsilon^{a_{0}...a_{p}} p.\xi_{2} p.\xi_{3}
 C_{ a_{0}...a_{p}}\bigg)\nonumber\\
+  k_{3c} \epsilon^{a_{0}...a_{p-1}c}  \xi_{2i}\xi_{3j} \bigg(p^i C_{j a_{0}...a_{p-1}}-p^j C_{i a_{0}...a_{p-1}}\bigg)\bigg\}
 \label{lop33}\eeqa
 First we would like to reconstruct the bulk poles that are mentioned in the first 
 line of \reef{lop33}, where we need to actually consider the following sub-amplitude in an effective field theory 

\beqa
{\cal A}&=&V^i_{\alpha}(C_{p+1},\phi_3,\phi)G^{ij}_{\alpha\beta}(\phi)V^j_{\beta}(\phi,A_1,\phi_2),\labell{amp44390}
\eeqa

$V^j_{\beta}(\phi,A_1,\phi_2)$  must be re-constructed by means of  the standard scalar fields's kinetic term in 
DBI action that  has no correction and has already been fixed in the effective action as  $\frac{(2\pi\alpha')^2}{2}  \Tr(D_a\phi^i D^a\phi_i)$ and the other vertices are

\beqa
V^j_{\beta}(\phi,A_1,\phi_2)&=&-2i (2\pi\alpha')^2T_p k_2.\xi_1
\xi^j_2 \Tr(\lambda_1\lambda_2\lambda_\beta)
\nonumber\\
G_{\alpha\beta}^{ij}(\phi)&=&\frac{-i}{(2\pi\alpha')^2T_p}\frac{\delta^{ij}
\delta_{\alpha\beta}}{k^2}\,\,\, ,
\label{ver138}
\eeqa

 so that $k^2=-(k_2+k_1)^2=t$  is replaced in the propagator. 
 
 \vskip.1in

To explore the vertex of $V^i_{\alpha}(C_{p+1},\phi_3,\phi)$ at leading order, one needs to keep in mind the following vertex
$\frac{\mu_p(2\pi\alpha')^2}{2(p+1)!}\int d^{p+1}\sigma
\eps^{a_0\cdots a_{p}}\ \Tr(\phi^j\phi^i)
\prt_i \prt_j C_{a_0\cdots a_{p}}$  to be able to  extract the vertex of an on-shell scalar and an off-shell scalar 
field as well as the potential C-term as follows

\beqa
V^i_{\alpha}(C_{p+1},\phi_3,\phi)&=&\frac{\mu_p(2\pi\alpha')^2}{(p+1)!}
 p^i p.\xi_{3} \eps^{a_0\cdots a_{p}} C_{a_0\cdots a_{p}}\Tr(\lambda_3\lambda_{\alpha}) 
\label{ppo90}\eeqa
 
Substituting \reef{ppo90} and \reef{ver138} into \reef{amp44390}, we find out the first t-channel bulk singularity of string amplitude. The propagator is fixed and there is no correction to $V^j_{\beta}(\phi,A_1,\phi_2)$, given the fact that it is obtained from kinetic term, therefore we conclude that there is no way of producing all the other bulk t-channel singularities, except  one  inserts all order $\alpha'$ higher derivative corrections to the following coupling 

\beqa
&&\frac{\mu_p(2\pi\alpha')^2}{2(p+1)!}\int d^{p+1}\sigma
\eps^{a_0\cdots a_{p}} \sum_{n=-1}^{\infty} b_n (\alpha')^n  \bigg(\Tr(D_{a_{1}}...D_{a_{n}}\phi^iD^{a_{1}}...D^{a_{n}}\phi^j)\bigg)
\prt_i\prt_j C_{a_0\cdots a_{p}}\label{5gh}
\eeqa

to actually derive the all order extended  vertex operator of  $V^i_{\alpha}(C_{p+1},\phi_3,\phi)$ as below
  
  \beqa
V^i_{\alpha}(C_{p+1},\phi_3,\phi)&=&\frac{\mu_p(2\pi\alpha')^2}{(p+1)!}\Tr(\lambda_3\lambda_{\alpha})
\eps^{a_0\cdots a_{p}}\sum_{n=-1}^{\infty} b_n (k_3.k)^{n+1} p.\xi_{3} p^i C_{a_0\cdots a_{p}}\label{ppo291}\eeqa
    where  $\sum_{n=-1}^{\infty} b_n (k_3.k)^{n+1}=\sum_{n=-1}^{\infty} b_n (s+u)^{n+1}$ should be used. 
    Now if one inserts \reef{ppo291}  into \reef{amp44390} and keeps fixed \reef{ver138} then all order
    t-channel bulk singularities  in the EFT are found out to be
    
 \beqa
 \sum_{n=-1}^{\infty}b_n  \frac{1}{t}(u+s)^{n+1} \frac{16\pi^2 \mu_p}{(p+1)!}  2k_2.\xi_1 \eps^{a_{0}\cdots a_{p}}C_{a_{0}\cdots a_{p}} p.\xi_{2} p.\xi_3
\label{tpoles2}\eeqa

which are precisely all the t-channel bulk poles  of the string amplitude  that appeared in the first line of \reef{lop33},
so we could regenerate them in an EFT as promised.

\vskip.1in

Eventually we would like to produce the rest of the t-channel bulk poles as follows 

 \beqa
 2i k_2.\xi_{1}    \frac{16 \pi^2 \mu_p}{(p+1)!} \sum_{n=-1}^{\infty}b_n\frac{1}{t}(u+s)^{n+1} \Tr(\lambda_1\lambda_2\lambda_3) k_{3c} \epsilon^{a_{0}...a_{p-1}c}  \xi_{2i}\xi_{3j} \bigg(p^i C_{j a_{0}...a_{p-1}}-p^j C_{i a_{0}...a_{p-1}}\bigg) \label{lop33678}\eeqa
 
 where the same field theory amplitude   \reef{amp44390}  is needed.
 Note that  due to $(-p^j C_{i a_{0}...a_{p-1}})$ term in string amplitude, one might think that we just
 need to employ one scalar field from Taylor expansion and the other external scalar field from Pull-Back of brane in an EFT but
 as we can see from string amplitude we need to produce the other term $(p^i C_{j a_{0}...a_{p-1}})$ in an EFT as well, so that the proper combination of terms in EFT is needed.

 \vskip.1in
 
 Suppose both external scalar fields come from   pull-back of brane  as
 \beqa
\frac{(2\pi\alpha')^2\mu_p}{2}\int d^{p+1}\sigma \frac{1}{(p-1)!}
\eps^{a_0\cdots a_{p}}\Tr\left(D_{a_0}\phi^i\,D_{a_1}\phi^j\right)
C_{ija_2\cdots a_{p}}\label{220}\eeqa

More significantly, consider the following   Bianchi identity 
 \beqa
 \eps^{a_0\cdots a_{p}} \bigg( -p_{a_{p}} (p+1) H^{ij}_{a_0\cdots a_{p-1}}-p^j H^{i}_{a_0\cdots a_{p}}+p^i H^{j}_{a_0\cdots a_{p}}\bigg)&=&d H^{p+2}=0
 \label{BI12}\eeqa
  extract the momentum of RR to make it just in terms of the potential of RR  as below 
 \beqa
   p_{a_{0}} \eps^{a_0\cdots a_{p}} \bigg( -p_{a_{p}} p (p+1) C_{ija_1\cdots a_{p-1}}-p^j
   C_{ia_1\cdots a_{p}}+p^i C_{ja_1\cdots a_{p}}\bigg)&=&0\label{mm11}\eeqa

Now if we extract the vertex of an on-shell scalar, an off-shell scalar field and a potential C- field from \reef{220} and
bear in mind the fact that the covariant derivative $D_{a_0}$ can act just on C-field ( also taking integration by parts), 
expecting to obtain the following vertex operator 
\beqa
V_{i\alpha}(C_{p+1},\phi_3,\phi)&=&\frac{\mu_p(2\pi\alpha')^2 p (p+1)}{(p+1)!}\Tr(\lambda_3\lambda_{\alpha})
\eps^{a_0\cdots a_{p}}  C_{ija_1\cdots a_{p-1}}k_{3a_{0}} p_{pa_{p}}\xi_{3j} 
\label{ppo9632}\eeqa

Indeed we now can use \reef{mm11} to be able to replace in \reef{ppo9632} $ p_{a_{p}} p (p+1) C_{ija_1\cdots a_{p-1}}$ in 
terms of $(-p^j C_{ia_1\cdots a_{p}}+p^i C_{ja_1\cdots a_{p}})$. By doing so and taking into account \reef{ver138}, replacing
\reef{mm11} inside \reef{ppo9632} as well as holding \reef{amp44390}, we are able to construct out just the first t-channel bulk 
singularity structure of \reef{lop33678} in an EFT. 

\vskip.1in

Given the previous clarifications and in order to regenerate  an infinite number of  t- channel bulk singularity structures in EFT,
one has to apply the correct  higher derivative corrections to pull-back as follows 

  \beqa
\frac{(2\pi\alpha')^2\mu_p}{2}\int d^{p+1}\sigma \frac{1}{(p-1)!}
\eps^{a_0\cdots a_{p}}     \sum_{n=-1}^{\infty} b_n (\alpha')^{n}  \Tr\left(D_{a_0}D_{a_{1}}...D _{a_{n}}\phi^i\,D_{a_1}D^{a_{1}}...D^{a_{n}}\phi^j\right)
C_{ija_2\cdots a_{p}}\label{220032}\eeqa

 so that the all order extension of the above vertex operator would be gained as follows
 
 \beqa
V_{i\alpha}(C_{p+1},\phi_3,\phi)= \Tr(\lambda_3\lambda_{\alpha})\frac{\mu_p(2\pi\alpha')^2 }{(p-1)!}
\eps^{a_0\cdots a_{p}} \sum_{n=-1}^{\infty} b_n (s+u)^{n+1} C_{ija_1\cdots a_{p-1}}k_{3a_{0}} p_{pa_{p}}\xi_{3j} 
\label{ppo963244}\eeqa

 Once more the contributions of  \reef{ver138}, replacement 
\reef{mm11} inside \reef{ppo963244} as well as the sub field theory amplitude \reef{amp44390},
 are taken. Having carried it out, we would be able to precisely produce all order t-channel bulk 
singularity structures of \reef{lop33678} in an effective field theory as well.

This ends our goal of producing an infinite number of t,s-channel bulk singularity structures of BPS branes.
It is worth mentioning that, there is another way of producing t,s-channel bulk poles in such a way that one needs to 
relate combination of certain terms in the effective actions of BPS branes, lets devote the rest of this section to it.

Consider the action where an scalar comes from Taylor expansion and the other scalar comes from pull-back   as follows 
  \beqa
(2\pi\alpha')^2\mu_p \int d^{p+1}\sigma \frac{1}{(p)!}
\eps^{a_0\cdots a_{p}}
\Tr\left(\phi^j D_{a_0}\phi^i\right)
\prt_jC_{ia_1\cdots a_{p}}
\label{221}
\eeqa
and add \reef{220} with \reef{221} as well as the terms  that have the same order in $\alpha'$  such as  Myers terms 
\beqa
{i\over4}(2\pi\alpha')^2\mu_p\int d^{p+1}\sigma {1\over(p-1)!} \eps^{a_0\cdots a_{p}}
\,\Tr\left(F_{a_0a_1}[\phi^j,\phi^i]\right)C_{ija_2\cdots a_{p}} .
\label{6733}
\eeqa

so that after having taken into account the integrations by parts, we would have left  with the desired action as 

\beqa
(2\pi\alpha')^2 \mu_p\int d^{p+1}\sigma {1\over p!}
\eps^{a_0\cdots a_{p}} \Tr\left(D_{a_0}\phi^j\phi^i\right)
p^i C_{ja_1\cdots a_{p}}
\label{2245}\eeqa

One may use \reef{2245} to extract $V^i_{\alpha}(C_{p+1},\phi_3,\phi)$ as  
\beqa
V^i_{\alpha}(C_{p+1},\phi_3,\phi)&=&\frac{\mu_p(2\pi\alpha')^2}{p!}\Tr(\lambda_3\lambda_{\alpha})
\eps^{a_0\cdots a_{p}} p^i C_{ja_1\cdots a_{p}}k_{3a_{0}} \xi_{3j} 
\label{ppo96}\eeqa

Furthermore, we 
 might consider the other term  so that this turn $\phi^i$ comes from pull-back and $\phi^j$ comes from  Taylor expansion as follows 
\beqa
-(2\pi\alpha')^2 \mu_p\int d^{p+1}\sigma {1\over p!}
\eps^{a_0\cdots a_{p}} \Tr\left(D_{a_0}\phi^i\phi^j\right)
p^j C_{ia_1\cdots a_{p}}
\label{2246}\eeqa
keeping in mind  \reef{2246} and extracting the rest of the terms, $V^i_{\alpha}(C_{p+1},\phi_3,\phi)$ vertex is got to be   

\beqa
V^i_{\alpha}(C_{p+1},\phi_3,\phi)&=&-\frac{\mu_p(2\pi\alpha')^2}{p!}\Tr(\lambda_3\lambda_{\alpha})
\eps^{a_0\cdots a_{p}} p^j C_{ia_1\cdots a_{p}}(k_3+p)_{a_{0}} \xi_{3j} 
\label{ppo966}\eeqa

Now we may want to add \reef{ppo96} with \reef{ppo966} and use the Bianchi identity $  p_{a_{0}} \eps^{a_0\cdots a_{p}}=0 $ 
to produce the leading order of the vertex operator in an EFT so that  the first t-channel bulk singularity structure of 
\reef{lop33678} is produced. We could apply the correct higher derivative corrections to  \reef{2245} and \reef{2246} such as  
\beqa
&&\frac{\mu_p(2\pi\alpha')^2}{p!}\int d^{p+1}\sigma
\eps^{a_0\cdots a_{p}} \sum_{n=-1}^{\infty} b_n (\alpha')^{n} \Tr\left(D_{a_{0}} D_{a_{1}}...D _{a_{n}}\phi^j
D^{a_{1}}...D^{a_{n}}\phi^i\right)
p^i C_{ja_1\cdots a_{p}}
\label{5gh}
\eeqa
 to indeed get to all  order t,s-channel bulk poles in an EFT. Finally, Let us turn to all contact interactions as well.

 \section{ All order $\alpha'$ Contact Interaction Analysis
}

Let us construct the complete and all order contact terms of this S-matrix.
If we  start to add all the first terms of ${\cal A}_{82},{\cal A}_{83},{\cal A}_{84}$ of asymmetric amplitude in
\reef{711u} we then derive
\beqa
 i sut \Tr(P_{-}\fsC_{(n-1)}M_p \Gamma^{jiba})\xi_{2i}\xi_{3j}   \xi_{1a} (k_{3}+k_{2} +k_{1})_{b} L_1 \eeqa
 where by using momentum conservation along the brane , the above terms exactly produce ${\cal A}_{3}$ 
 of symmetric amplitude in  \reef{11u89}.
 
 \vskip.1in

Considering the 1st terms ${\cal A}_{42},{\cal A}_{72}$ and extracting the trace , one finds out the following terms

\beqa
-iust L_1   \xi_{1a} \xi_{2i} \xi_{3j} \frac{16}{(p+1)!} \epsilon^{a_{0}...a_{p-1}} (p^i C_{j a_{0}...a_{p-1}}-p^j C_{i a_{0}...a_{p-1}})
\eeqa

where we consider these new terms later on.

 
\vskip.1in

 Note that ${\cal A}_{81}$
precisely does produce the 1st term of  ${\cal A}_{1}$ of symmetric result of \reef{11u89}. On the other hand,
using momentum conservation, ${\cal A}_{71}$  can be written down as

\beqa
i p.\xi_2 \Tr(P_{-}\fsC_{(n-1)}M_p \Gamma^{jcad})\xi_{1a}\xi_{3j}  k_{3c} L_4 (-p-k_3-k_2)_{d}\label{lop98}\eeqa
where using the anti symmetric property of $\epsilon$ , one reveals that the 2nd term in \reef{lop98} 
has no contribution. Making use of $(p\fsC=\fsH)$, the 1st term of \reef{lop98} generates the 3rd term of ${\cal A}_{1}$  
of symmetric amplitude in  \reef{11u89} (which is contact interaction term), while the last term in \reef{lop98} 
is an extra contact interaction that we take it into account  in a moment. Likewise,  the same analysis holds for
${\cal A}_{41}$ as follows
\beqa
i p.\xi_3 \Tr(P_{-}\fsC_{(n-1)}M_p \Gamma^{ibad})\xi_{1a}\xi_{2i}  k_{2b} L_4 (-p-k_2-k_3)_{d}\label{lop989}\eeqa
where the 1st term in \reef{lop989} does reconstruct the 2nd term of ${\cal A}_{1}$ 
of symmetric amplitude in  \reef{11u89} ( its second term has zero contribution) 
, meanwhile the last term in \reef{lop989} is an extra contact interaction that we regard it in the next 
sections as well.

\vskip.1in

Ultimately, the 2nd term  ${\cal A}_{3}$ of asymmetric amplitude in \reef{711u} 
is written down as 
\beqa
-i p.\xi_3 p.\xi_2 \Tr(P_{-}\fsC_{(n-1)}M_p \Gamma^{ad})\xi_{1a} L_4 (-p-k_2-k_3)_{d}\label{lop9898}\eeqa
indeed the first term in above equation regenerates the 4th contact term ${\cal A}_{1}$  of symmetric amplitude
in  \reef{11u89}.

\vskip.1in

Hence,  we are able to produce all the contact interactions of \reef{11u89} by using the elements of
asymmetric S-matrix. However, the last two terms of 
\reef{lop9898} are extra contact terms in asymmetric S-matrix \reef{711u} and we claim their contribution is needed to 
our actual S-matrix as we demonstrate it right now.

\vskip.1in

Let us just end this section by adding all the extra contact interactions, extracting all the traces  and using
the antisymmetric property of $\epsilon$ tensor  to be able to essentially obtain the following new contact terms to all orders
\beqa
i \xi_{1a}\xi_{2i}\xi_{3j}     \frac{16}{(p-1)!} \bigg\{L_4 \bigg((k_2+k_3)_{d} p^i p^j \epsilon^{a_{0}...a_{p-2}ad}  C_{ a_{0}...a_{p-2}}\nonumber\\
+ k_{3d} k_{2b} \epsilon^{a_{0}...a_{p-3}bad} (p^i C_{j a_{0}...a_{p-3}}-p^j C_{i a_{0}...a_{p-3}})\bigg)\nonumber\\
-ust   L_1  \frac{1}{p (p+1)} \epsilon^{a_{0}...a_{p-1}a} \bigg( p^i C_{j a_{0}...a_{p-1}}-p^j C_{i a_{0}...a_{p-1}}\bigg)\bigg\}
 \label{lop0055}\eeqa


The first term  in \reef{lop0055} is symmetric under interchanging 
both scalar fields and is needed in the string theory amplitude as it can be explored  by means of Taylor expansions 
of the Effective field theory couplings, whereas its infinite higher derivative corrections can also be explored by 
applying appropriate higher derivative corrections to either Wess-Zumino or Chern-Simons couplings. As the method for extracting
all order $\alpha'$ corrections to BPS contact interactions has been comprehensively explained in section five of 
\cite{Hatefi:2012zh} and 
\cite{Hatefi:2015ora} accordingly. Note also, as we explained earlier on, by combining  the couplings of \reef{2245} and \reef{2246}
 and inserting the correct higher derivative corrections to them, one immediately starts to generate all order $\alpha'$ corrections 
 to all new BPS contact terms that have been released in \reef{lop0055}.

 Notice that, since we have 
 found all these terms by direct S-matrix analysis , one assured that the coefficients 
 of the corrections are also exact and have no ambiguity any more. Ultimately, it is worth to point out that several
 new couplings within new structures have also been explored in section 9 of \cite{Hatefi:2015ora}.

 \vskip.1in

By explicit computations, it was also revealed that in an effective field theory, most of the super gravity field contents in
the actions should be various functions of SYM. Because it is evidently realized that either Taylor expanded of transverse scalar
fields (for the background
fields) or some combinations of pull-back , Taylor expansion employed and this has been first regarded in 
the so called Dielectric effect \cite{Myers:1999ps}.

 \vskip.1in
 
 One may have some hopes in figuring out the importance of 
 the above new couplings, results for the S-matrices to construct not only  future research areas in theoretical high energy physics , most
 notably in D-branes area but also in discovering new sort of Myers terms as well as constructing higher point functions or 
 mathematical results (symmetries) behind the 
scattering amplitude prospectives. We intend to investigate and go through some of unanswered open questions in near future.

\section{Conclusion}
In this paper we started exploring the complete form of the S-matrix of two transverse scalar fields, a gauge field and a potential RR 
form-field in type IIA,IIB superstring theory, namely among other contents, we have derived even the terms that explicitly carry 
$p.\xi_1$ and  $p.\xi_2$ elements in the string amplitude. For the last part of the S-matrix we needed to find out the explicit form
of integrations on upper half plane for 
arbitrary combinations of Mandelstam variables including the terms that do clealy involve 
$\int d^2z |1-z|^{a} |z|^{b} (z - \bar{z})^{c}
(z + \bar{z})^{3}$, where this was derived.

We also generated an infinite number of $t,s$ channel bulk singularity structures by means of all order $\alpha'$ corrections to
pull-back of brane and highlighted the fact that unlike  
\cite{Hatefi:2016enc}, neither there are $u$- channel bulk nor $(t+s+u)$- channel bulk singularity structures.
Due to presence of the complete form of S-matrix, several new contact interaction couplings in
\reef{lop0055} have been discovered, whereas these terms
can be verified at the level of effective field theory by either the combinations of Myers terms, pull-back, Taylor expanded of 
scalar fields or the mixed combination of the couplings of \reef{2245} and \reef{2246}.  
Given the method that is explained in section five of 
\cite{Hatefi:2012zh} and \cite{Hatefi:2015ora}, one is able to constantly apply the higher derivative 
corrections on contact terms and immediately explores their generalization to all orders in $\alpha'$.
\section*{Acknowledgements}
The author would like to thank P. Anastasopoulos, N.Arkani-Hamed, A. Brandhuber, M.Douglas, C.Hull, W. Lerche, R.Myers, R.Russo,
and H. Steinacker for very valuable discussions and comments. Parts of the computations of this paper 
were taken place during my second post doctoral research at Queen Mary University of London and I indeed thank QMUL
 for the hospitality while this work was being completed. This work was supported by the FWF project P26731-N27.  
  

\end{document}